\documentstyle[12pt]{article}

\newcommand{\be}{\begin{equation}}
\newcommand{\ee}{\end{equation}}
\newcommand{\bes}{\begin{eqnarray}}
\newcommand{\ees}{\end{eqnarray}}
\newcommand{\bea}{\begin{equation}\begin{array}{lcl}}
\newcommand{\eea}[1]{\end{array}\label{#1}\end{equation}}
\newcommand{\ba}{\begin{array}}
\newcommand{\ea}{\end{array}}

\newcommand{\refe}[1]{(\ref{#1})}
\newcommand{\ns}{\normalsize}
\newcommand{\ler}{\stackrel{\scriptstyle <}{\scriptstyle\sim}}
\begin{document}
\begin{titlepage}
\setcounter{page}{1}

\title{{\LARGE Discrete Gauge Symmetries in Axionic Extensions of the SSM}
         \thanks{supported by Deutsche Forschungsgemeinschaft}}
\author{{\large E.~J.~Chun\thanks{supported by a KOSEF fellowship}
        \hspace{0.02cm} and
        A.~Lukas}\\[0.8cm]
        {\ns Physik Department}\\
        {\ns Technische Universit\"at M\"unchen}\\
        {\ns D-8046 Garching, Germany}\\
        {\ns and}\\[0.1cm]
        {\ns Max-Planck-Institut f\"ur Physik}\\
        {\ns Werner-Heisenberg-Institut}\\
        {\ns P.~O.~Box 40 12 12, Munich, Germany}}

\date{{\ns August 1992}}

\maketitle

\setlength{\unitlength}{1cm}
\begin{picture}(5,1)(-12.0,-16.0)
\put(0,-0.5){TUM - TH - 150/92}
\put(0,-1.0){MPI - Ph/92 - 71}
\end{picture}

\begin{abstract}
We examine discrete gauge symmetries in axionic extensions of the SSM
which provide a solution of the $\mu$-problem. Automatic-PQ symmetry
and proton stability are shown to be guaranteed by certain discrete
symmetries. Focusing on the L-violating discrete symmetries we discuss
two sources of neutrino masses and their relevance for the solar neutrino
problem.
\end{abstract}

\thispagestyle{empty}
\end{titlepage}
\clearpage

\setcounter{page}{1}
In the conventional standard model baryon (B) and lepton number
(L) invariance are automatic consequences of the gauge symmetry and
the particle contents. Contrary to this the supersymmetric standard model
(SSM) has to be endowed with certain discrete symmetries (typically R parity)
to enforce the B- and/or L-invariance.
An interesting possibility is to interpret these discrete symmetries
as gauge symmetries~\cite{kw} which can be remnants of a broken unifying
gauge group. Moreover, discrete gauge symmetries can stabilize
B- and L-invariance even if quantum gravity effects violate global symmetries.
The discrete symmetries should then fulfill certain anomaly-free
conditions~\cite{bd,ibanez_ross}.
Recently, it has been found that proton stability or solar neutrino
mixing can be consistent with certain classes of anomaly-free discrete
symmetries~\cite{ibanez_ross,kap_mayr_nilles}.

In this paper, we address the question of an ``automatic''
PQ-symmetry~\cite{kim_physrep} in the
SSM guaranteed by a certain discrete gauge symmetry which is
assigned to prevent serious B- or L-violation. As a consequence all desirable
global symmetries of the model can be made stable against possible
quantum gravity effects.

As we will see it appears to be very difficult to implement an
automatic PQ-symmetry of the Kim-type~\cite{kim_axion}.
Therefore we concentrate on the Dine-Fishler-Srednicki-axion~\cite{dfs}
and we extend the Higgs sector of the SSM
by adding some standard model singlets $A_i$ (PQ-fields).
The requirements of automatic PQ lead to a solution of the
$\mu$--problem~\cite{kim_nilles,kim} by introducing the $d=5$ operator
$H \bar H A^2$. When the field $A$ gets a vacuum expectation value (VEV)
of the order of $f_a$ ($f_a$ is the PQ-symmetry breaking scale with
$10^{10}$ GeV $<f_a<10^{12}$ GeV) the generated $\mu$-term receives
the correct order of magnitude.

Depending on the PQ-field contents and discrete symmetries,
various kinds of B-, L- or PQ-violating operators appear.
One should check whether some combinations of these operators can be eliminated
to ensure proton stability and the axion solution of the strong CP problem.
In particular, the axion solution is very sensitive to the presence of higher
dimensional PQ-nonsinglets  because of the small axion
mass~\cite{automatic_PQ}.

The fermionic part of the chiral superfields $A_i$ can serve
as a right handed neutrino (RHN) due to the possible couplings $L\bar{H}A$
and $L\bar{H}A^2$.
Clearly neutrino masses can be generated by nontrivial VEV's $<A>$.
In addition, the see-saw mechanism~\cite{see_saw} can be operative depending
on the masses of the fields $A$. Indeed this amounts to the supersymmetric
version of the neutrino-axion connection investigated in~\cite{n_a_conn}.
Another purpose of this paper is to
discuss the phenomenological implications for
neutrino physics in this context. Especially we address to possible
solutions of the solar-neutrino-problem recently confirmed as a
problem of elementary particle physics by the
measurement of the GALLEX-Collaboration~\cite{GALLEX,bludman}.\\

 We begin our discussion with the DFS-type axionic extensions~\cite{dfs}
of the SSM. The classification of discrete gauge symmetries for the
SSM was given by Ib\'a\~nez and Ross~\cite{ibanez_ross}.
They wrote all discrete symmetries $g_N$
compatible with ``standard'' operators in terms of three generators~:
\be
  g_N=R^mA^nL^l \label{standard_symm}
\ee
which leads to charges of the standard particles as given in
table~\ref{tab : disc_charges}.
\begin{table}[h]
 \begin{center}
 \begin{tabular}{|c|c|c|c|c|c|c|c|}
  \hline
   &$Q$ & $u$ & $d$ & $L$ & $e$ & $H$ & $\bar{H}$  \\ \hline
   $\alpha$ & $0$ & $-m$ & $m-n$ & $-n+l$ & $m-l$ & $-m+n$ & $m$  \\ \hline
 \end{tabular}
 \end{center}
 \caption{Discrete charges of standard particles.}
 \label{tab : disc_charges}
\end{table}
Additionally this symmetries have to fulfill anomaly cancelation
conditions due to the gravity-, $SU(3)$- and $SU(2)$-interaction. In our
extension the generators of eq.~\refe{standard_symm} have to be
continued to the new fields $A_i$. Moreover new generators can appear
which only act non trivially on the $A_i$. What happens in detail of course
depends on the explicit form of the superpotential.
Some examples will be given later.
The modified form of the anomaly cancelation conditions reads
\bes
 3m + n + k &=& {1\over2}\eta s N \quad {\rm for\; gravity} \nonumber\\
 3n &=& 0  \quad\quad\quad {\rm for}\; SU(3) \label{anomalies} \\
 3l + n &=& 0 \quad\quad\quad {\rm for}\; SU(2), \nonumber
\ees
where $\eta=1$ for $N$ even and $\eta=0$ for $N$ odd.
Since we added SM-singlets only the gravity anomaly is modified by
$k=\sum \alpha(A_i)$.
Parts of our discussion can be done referring to the value of $k$ instead
of using the explicit form of $W(A_i)$.

We start with the implications of the automatic-PQ idea.
An explicit $H\bar{H}$-term violates
the PQ-symmetry and so we should demand $n\neq 0$. The last two anomaly
free conditions show that only a restricted class of discrete symmetries
is compatible with this requirement~:
\be
  N = 9r,\quad  n =  3r, \quad l = 2r, \quad r=1,2,3,... \label{symm}
\ee
Having forbidden $H\bar{H}$ we are forced to generate this term
dynamically. This amounts to solving the $\mu$-problem.
Introducing the conventional DFS-coupling $H\bar{H}A_1$,however, needs
a very small coupling constant. Therefore we use one of the terms
$H\bar{H}A_1A_2$ or $H\bar{H}A_1^2$ which naturally lead to the desired
order of magnitude. For discrete symmetries with $N=3^i$
the second term cannot be used because the coupling $A_1^3$ then violates
PQ-symmetry. Moreover PQ-symmetry has to be respected by higher dimensional
operators. As shown recently~\cite{automatic_PQ}, the lowest PQ-nonsinglet
operators constructed out of only PQ-fields should be of
(component field) dimension $d=10$.
But the restriction is less stringent for the standard fields
since they have much smaller VEV's  than PQ-fields.
For standard operators, it is enough to forbid
$L\bar{H}$ \footnote{This is also necessary in the
spirit of the solution to the $\mu$-problem.}, i.~e.~$m\neq r$, to protect
the axion solution.
For mixed operators (operators containing standard fields and fields $A_i$)
only the combination of $L\bar{H}A_i$ and a PQ-nonsinglet $L\bar{H}A_jA_k$
can cause a problem.
Of course this depends on the VEV's of the $A$'s.
As we will see from the examples, it is possible to forbid either of these two
if we do not want to introduce too many fields.
Dangerous $d=7$ operators generally appear in the pure $A_i$-sector.
To make a concrete discussion, let us take a $\mu$-term $H\bar{H}A_1^2$ and a
superpotential $W =(A_1 A_2 - f_a^2)Y$.
As a consequence of the second anomaly-free condition in
eq.~\refe{anomalies}, one finds the allowed operators~:
$A_i^6$ or $A_1^5 A_2^*$, etc.
Now we have to worry about the PQ-nonsinglet operators
in the scalar potential like $ a_1^5 a_2^* y^* $.
It can be obtained from the terms $A_1 A_2 Y$ and $A_1^6$ in the
superpotential. At first glance, it does not make a problem since
$<Y> = 0$ in the global SUSY case.
But if we take account of the full supergravity Lagrangian,
the field $Y$ can develop a VEV of the order of $f^{p+1}/M_{\rm pl}^p$.
Here $p$ can depend on the specific form of the
superpotential~\cite{ch_ki_ni}.
For instance, the above superpotential gives $p = 1$.
So, {\it one has to take a superpotential which gives $p \geq 3$ }
to protect the axion solution.
{}From this, we favor a very light axino (mass $\leq  10^{-4} eV$) since
its mass is proportional to $<Y>$.

 Now we turn to B-, L-number.
The discrete gauge symmetries can be classified according to the allowed
pattern of the B-, L-violating standard terms
$LQd,LLe$ and $udd$~\cite{ibanez_ross}~:
\bes
 {\rm GMP}&:&m\neq 4r,6r \nonumber \\
 {\rm GLP}&:&m=6r\\
 {\rm GBP}&:&m=4r\; . \nonumber
\ees
Proton decay safety on higher dimensional standard operators is ensured
if $QQQL,uude$ and the combinations $(udd,QueH/QuL^*/ud^*e)$
and $(LQd,QQQH)$ are forbidden.
{\em This is fulfilled for all our symmetries in eq.~\refe{symm} !}
Additionally new B-, L-violating operators can appear by adding singlet
fields $A_i$. The combinations $(LQdA_i^p,uddA_j^q)$ with $p+q\leq 3$
should be nonsinglets under discrete symmetry to avoid fast proton decay.\\
Interesting patterns of allowed and forbidden B- and L-violating
operators with dimension $\leq 5$ appear if all
charges $\alpha (A_i)$ are multiples of 3.
This is always the case if the coupling $H\bar{H}A_1^2$
is demanded to be present. But also for $H\bar{H}A_1A_2$
there exists a class of superpotentials with this property.
We define $c=m$ mod 3. Then for $r$ mod $3\neq 0$ three patterns of
operators occur two of them automatically lead to a proton decay safe
theory (see table~\ref{tab : op_patterns}).
\begin{table}
 \begin{center}
 \begin{tabular}{|c|c|c||c|}
  \hline
  &forbidden&allowed for some $m$&$k$-values for $N=18$\\ \hline
  $c=0\; {\rm mod}\; 3$&L-violating&B-violating&$3, -6$\\ \hline
  $c=r\; {\rm mod}\; 3$&B-violating&L-violating&$6, -3$\\ \hline
  $c=2r\; {\rm mod}\; 3$&all&none&$0,9$\\ \hline
 \end{tabular}
 \end{center}
 \caption{Patterns of allowed and forbidden operators for different
          classes of potentials $W(A_i)$.}
 \label{tab : op_patterns}
\end{table}
For $r$ mod $3=0$ all operators can be allowed in the case $c=0$ whereas
in the opposite case all operators are forbidden. The $k$-values in terms
of the class $c$ are given by
\be
 \frac{k}{3}\;{\rm mod}\;3=\left\{ \ba{ll}
                       \frac{r}{2}-c&r\;{\rm even}\\
                       -r-c&r\;{\rm odd} \ea \right. . \label{k_values}
\ee

 We saw that certain discrete symmetries can allow L-violation. In this class
of symmetries neutrino masses can be generated with the fields $A_i$
serving as RHN's. The relevant couplings are the following~:
\[
 a_{ij}L_i\bar{H}A_j+\frac{b_{ijk}}{M}L_i\bar{H}A_jA_k+
 \frac{c_{ij}}{2M}H\bar{H}A_iA_j+
\]
\[
 \frac{m_{ij}}{2}A_iA_j+\frac{g_{ijk}}{3}A_iA_jA_k+\frac{d_{ij}}{2M}
 L_i\bar{H}L_j\bar{H}+\frac{k_i}{M}L_i\bar{H}H\bar{H}.
\]
The last two terms - if present - can produce neutrino masses $O(10^{-5})$
eV (the last term only one mass of this order). With this in mind we neglect
them in the following diagonalization procedure. The mass matrix of
$f=(\nu_i,\tilde{h},\tilde{\bar{h}},\tilde{a_k})$ can be read off from the
above couplings~:
\be
 M=\left( \ba{cccc} 0&0&c&M_D\\
                    0&0&\mu&b^T\\
                    c^T&\mu&0&a^T\\
                    M_D^T&b&a&M_M \ea \right).
\ee
Neglecting again contributions $O(10^{-5})$ eV one can diagonalize this
matrix in three steps~: rotating away the $\nu-\tilde{\bar{h}}$-mixing,
diagonalizing $M_M$ and applying see-saw mechanism for all ``large''
eigenvalues of $M_M$. Then we arrive at~:
\be
 \hat{M}=\left( \ba{cccccc}
        D_L\hat{L}^{-1}D_L^T+D_S\hat{S}^{-1}D_S^T&0&0&D_A&0&0\\
        0&0&\mu'&0&0&0\\
        0&\mu'&0&0&0&0\\
        D_A&0&0&\hat{A}&0&0\\
        0&0&0&0&\hat{S}&0\\
        0&0&0&0&0&\hat{L} \ea \right). \label{diag_matrix}
\ee
The diagonalized matrix $M_M$ has been divided into three parts~:
the axion and very small eigenvalue ($\ll m_{3/2}$) part $\hat{A}$,
the small eigenvalue part $\hat{S}$ with eigenvalues $\ler m_{3/2}$
and the large eigenvalue part $\hat{L}$ with
eigenvalues $O(f_a)$. The other matrices have been divided into
the corresponding parts. The equations show that the couplings
$L_i\bar{H}A_j$ may
only be present if the VEV of $A_j$ is small ($\ler O(m_{3/2})$) and $A_j$
does not mix with the light eigenmodes in the
$\hat{S}$- and $\hat{A}$-sector.
Under these assumptions the matrices $D_A$ and $D_S$ are at most of the
order $f_am_{3/2}/M\sim$ keV. The application of the see-saw
mechanism for $D_S$ is then justified.
{}From the matrix~\refe{diag_matrix} the following conclusions can be drawn~:
\begin{enumerate}
 \item In general $D_A$ will be nonzero if couplings
       $L\bar{H}AA$ are present. Then the axino mass determines the mass
       of one neutrino mode. For a light axino mass ($\ll$ keV)
       a keV-Dirac-neutrino $(\nu,\tilde{a})$ appears. Though there
       are some hints for a 17-keV-neutrino experimental
       results~\cite{17_keV_exp} as well as cosmological
       implications~\cite{17_keV_cosm} disfavor this possibility.
       For an axino mass in the keV-range a keV-Majorana-neutrino is present
       which is excluded by neutrino-less double beta decay.
       It remains the possibility of a heavy axino ($O(10^2$ GeV)) generating
       a Majorana neutrino mass of $O(10^{-3})$ eV. This is the correct order
       for the MSW-solution~\cite{msw} of the solar neutrino problem.
       However, the last possibility which is favorable from the viewpoint
       of neutrino physics is in danger to spoil the axion solution
       as discussed.
 \item Couplings $L\bar{H}A$ can be present under the above requirements.
       Together with the see-saw mechanism they lead to
       $O($eV) Majorana neutrinos (for couplings $O(1)$).
 \item Apart from these mechanisms neutrino masses relevant
       for the solar neutrino problem can be generated only
       by see-saw suppression through light
       fields $A_i$~: If $D_S$ is of $O(f_am_{3/2}/M)\sim$ keV then
       $A_i$ masses of $O(m_{3/2})$ lead to masses
       $O(f_a^2m_{3/2}/M^2)\sim 10^{-3}$ eV
       (for large $f_a$, $f_a\sim 10^{12}$ GeV). A possible source
       of a mass of $O(m_{3/2})$ can be radiative corrections~\cite{my}.
 \item The couplings $L\bar{H}L\bar{H}$ cause Majorana neutrino
       masses of $O(10^{-5}$ eV) in general for all flavors. The coupling
       $L\bar{H}H\bar{H}$ can generate only one mass of this order.
 \item Neutrinos with mass $O(eV)$ which can be candidates for Dark Matter
       can be achieved in two ways. According to the first remark the
       scale $f_a\sim 10^{12}$ GeV can combine with an axino mass
       ($\ler 1$ GeV). Alternatively the superpotential can allow for
       a see-saw suppression by a scale of $O(f_a)$ as in the second remark.
\end{enumerate}
{\em Remarkably solving the solar-neutrino-problem via masses as discussed
or via $LQd$~\cite{LQd} is only possible for the
second pattern in table~\ref{tab : op_patterns}. Proton decay safety
follows then automatically.}

Before going to the examples let us discuss the relation of the
automatic PQ-idea and the Kim-type realization of axion.
The essential coupling for the Kim-axion $K_1A_1K_2$ (where $K_1$
and $K_2$ are heavy quarks) leads to the following parameterization of
charges~: $\alpha(K_1)=-p+q$, $\alpha(K_2)=-p-q$, $\alpha(A_1)=2p$.
The anomaly cancelation conditions read
\bes
 3m + n + 2p +k &=& {1\over2}\eta s N  \nonumber \\
 3n + 2p &=& 0 \\
 3l + n &=& 0\; . \nonumber
\ees
A difficulty in implementing the Kim-axion is that the model is in danger
to possess an additional DFS-PQ-symmetry. Then the introduction of additional
quarks and moreover the whole Kim-PQ-mechanism is superfluous.
One possibility to avoid this is to forget about solving $\mu$-problem
and demand $H\bar{H}$ to be present. But in this case $n=0$ and the
anomaly free conditions imply
$\alpha(A_1)=0$. So there is no chance for automatic PQ. On the other
hand, if $n\neq 0$ at least one coupling $T$ which is invariant under
discrete $n$-symmetry but noninvariant under the corresponding global
$U(1)$ is required~:
$\alpha(T)=sn=s'N \quad (s\neq 0)$ and $\alpha_{U(1)}(T)\neq 0$.
Now for all small values of $s$ ($s=3/2,2,5/2..$) the anomaly
cancelation conditions show that some small power of $A_1$ is allowed.
But the generation of large $s$-values leads to complicated
potentials and the introduction of many fields $A_i$.
We conclude that an automatic pure Kim-PQ extension of the SSM cannot
be implemented in an easy way.\\

Now we discuss all acceptable potentials $W$ for $N=9, 18$
up to four fields $A_i$.
Later we will give one example for $N=36$ which naturally can allow three
couplings $L\bar{H}N_i$~\footnote{All PQ-fields with this kind of coupling
will be denoted by $N$ in the following.}.
The  superpotential $W$  should have the following general properties~:
The superpotential is the most general one compatible with its
discrete symmetry. Of course it should admit a PQ-symmetry.
The discrete charges $\alpha(A_i)$ should satisfy the constraint
$\sum \alpha(A_i)=k$.
To force the PQ-fields responsible for the $\mu$-term to get a nonvanishing
VEV the existence of a field $Y$ with $\alpha(Y)=0$ is necessary.
Finally, supersymmetry is supposed to be unbroken in the sector of singlet
fields $A_i$. One can assume the Polony form for the supersymmetry breaking.

Let us first consider the case with $N=9$. Here we have $r=1$.
We allow the coupling $H \bar{H} A_1 A_2$ assuming the discrete charges of
the $A$'s to be $\alpha(A_1, A_2) = (-1, -2)$
\footnote{Two other discrete charge assignments are possible~:
$\alpha(A_1, A_2) = (2, 4)$ or $(-4, 1)$.  The following arguments are also
applicable to these cases.}.
The anomaly free condition in eq.~\refe{anomalies} allows the solutions
$k =  3x$ for an integer $x$ given by $m = 2-x$ (mod 3).
We find that our minimal choice for the superpotential requires two more fields
with $\alpha(A_3, A_4) = (1, 2)$.
Dropping the cubic and quadratic term for $Y$, the corresponding superpotential
reads~:
\be
 W = y_1 A_1^2 A_4 + y_2 A_2 A_3^2 + (y_3 A_1 A_3 + y_4 A_2 A_4 - f^2)Y
 \label{z9_potential}
\ee
where $f \sim f_a$.  This corresponds to $k=0$.

{}From the previous discussion we know that certain combinations of
B- and L-violating operators should be avoided
($(LQdA^p,uddA^q)$ with $p+q\leq 3$, $L\bar{H}$, $L\bar{H}A$).
For any $m$ at least one of these operators is allowed.
Since we can hardly expect that introducing more fields can avoid this
difficulty, we exclude the possibility for $Z_9$ symmetry.

The next examples are the $Z_{18}$ symmetries with $r=2$.
The values of $k, m$ follow the pattern $k = 3x$, $m = 1 -x$ (mod 3) for
an integer $x$.
We now introduce a PQ-field $A_1$  with discrete charge $-3$ to allow the
coupling $H \bar H A_1^2$.  Automatic PQ-symmetry requires
$\alpha(A_i) \neq \pm 6$ for all PQ-fields.
In agreement with table~\ref{tab : op_patterns}, we find three classes
of symmetries which share the similar phenomenological property~:\\
(I) $m = 1$ mod 3; $ k= 0, 9$~:
The minimal choice of the PQ-fields  can be achieved by taking two fields $A_1,
A_2$ with $\alpha(A_1,A_2) = (-3, 3)$.
Then we take the superpotential $W = (\lambda A_1 A_2 - f^2) Y + m_N N^2$,
where $ f \simeq m_N \simeq f_a$.  The presence of the field $N$
is necessary to achieve $k=9$.
One finds that this class of discrete symmetries allows no $d=5$ operators.
Therefore, B- and L-symmetry are insured quite strongly.\\
(II)  $m = 0$ mod 3; $ k= 3, -6$~:
For this case we need three fields with $\alpha(A_1, A_2, A_3) = (-3, 3, 3)$
for $k= -3$ and one more field $N$ for $k=-6$.
The corresponding superpotential can be read off from eq.~\refe{three}
by exchanging $A_1$ and $A_3$.
As expected, this class  of discrete symmetries respects
L-invariance. But B-violation  can appear in renormalizable terms or
in the higher dimensional operators with PQ-fields. For instance, the
discrete symmetry allows $udd$, $uddN$, $Q^3 H$, $uddA_{2,3}$ and $uddA_1$
for $m=8,3,6,9$ and $-3$ respectively.\\
(III)  $m = 2$ mod 3; $ k= -3, 6$~:
In this case, we have three fields with
$\alpha(A_1, A_2, A_3) = (-3,-3, 3)$ for $k=-3$ and one more
field $N$ for $k= 6$.
One can build two superpotentials satisfying our general requirement,
\be \label{three}
  W = \left\{ \ba{l}
  (\lambda_1 A_1 A_3 +  \lambda_2 A_2 A_3 - f^2) Y  + m_N N^2 \\
  (\lambda_1 A_1 A_3 - f_1^2) Y_1 + (\lambda_2 A_2 A_3 - f_2^2) Y_2
  + m_N N^2 \\ \ea \right.\: .
  \ee
  The first superpotential has two  massless multiplets~:
  $\lambda_2 A_1 - \lambda_1 A_2$ and the axion multiplet.
  In the global SUSY case, the VEV of the first  multiplet
  is not determined. But it  is set to be 0 if a supergravity Lagrangian
  with minimal kinetic terms and a vanishing cosmological constant
  is considered.
  The second one gives  nontrivial VEV's to all the PQ fields
  and only one massless  (axion) multiplet is present.
  These features become important for neutrino physics.

Contrary to the above case,  B-invariance is now insured.
We discard the cases with $m = 2$ and $5, -1$ since they allow either
$L\bar{H}$ or $L\bar{H}A_i$.
The L-violating operators are $LQd$, $LLe$ and $L\bar{H}A_{1,2}^2$ for $m=8$;
$L\bar{H}A_3^2$ and $L\bar{H}H\bar{H}$ for $m=-4$;
$L\bar{H}N$, $LQdA_{1,2}$, $LLeA_{1,2}$ and $L\bar{H}L\bar{H}$ for $m=-7$.
According to the general discussion the neutrino masses in this three
cases behave as follows~: \\
(i) $m=8, -4$~: As  discussed before, the field $Y$ can
get a VEV of order of $10^2$ GeV to produce a heavy axino.
Then, there   may appear  one neutrino with mass $O(10^{-3})$ eV for the both
cases in eq.~\refe{three}) and one keV-Dirac neutrino for the first case only.
However, this  large $<Y>$ spoils the axion solution.
To avoid this,  we need a superpotential with
sufficiently small gravity-corrected VEV $<Y>$. This may result in two (one)
keV Dirac neutrino for the first (second) superpotential.
In this minimal example, we cannot, therefore,
achieve the desirable features for  neutrino masses.\\
(ii) $m=-7$~: If the field $N$ is present ($k=6$) it is the only one which
can couple to $L\bar{H}$. It leads to one Majorana neutrino with mass
$O($eV). Here we can take a small VEV $<Y>$ to protect the axion solution
without arriving at a keV-Dirac neutrino. Due to the allowed coupling
$L\bar{H}L\bar{H}$ the other neutrino modes will have in general
mass $O(10^{-5})$ eV.

Up to now, we discussed the simple case with one PQ-field coupled
to $H\bar{H}$. If the term $H\bar{H}A_1A_2$ is forced to be allowed we find
that - as in the $Z_9$-case - at least four PQ-fields are necessary
to build the potential. For $\alpha(A_1)=3,9,15$ all charges
$\alpha(A_i)$ are multiples of three and the classification of operator
patterns from the general part is applicable. We find four types of
minimal potentials which cover the three possible patterns of allowed
and forbidden operators. For $\alpha(A_1)\neq 3,9,15$ the only minimal
potential is given by eq.~\refe{z9_potential} with $Z_{18}$-symmetry.
Because of the extended symmetry it cannot be ruled out as in the $Z_9$-case.

We saw that the most promising way of generating neutrino masses without
spoiling the axion solution is to introduce couplings $L\bar{H}N$.
In eq.~\refe{three} the mass of the right-handed neutrino was
put in  by hand. Now we construct an example which shows a relation between
the axion scale and the right-handed neutrino scale.
For this, let us go to the $Z_{36}$ symmetry. We take PQ-fields whose
discrete charges are given by $\alpha(A_{1,2}, A_3, N_{1,2,3}) = (-6, 6, 3)$.
This allows us to build a superpotential $W = (\lambda_1 A_1A_3 + \lambda_2
A_2A_3 - f^2) Y + h_i A_3N_i^2 $ with $k=3$.
Now the discrete symmetry with
$m=1$ permits the desired coupling $L_i\bar{H}N_j$ and forbids any
$d=5$ operators as well as B- and L-violating $d=4$ operators.
One finds that the masses of right-handed neutrinos are given by
$h_i <A_3>$. Irrespectively of $<Y>$ the axino receives a
radiative-corrected  mass of $O(10^2)$ GeV by the coupling
$A_3 N_i^2$~\cite{my}.\\

 We finish our discussion with some remarks about generation dependent
 discrete symmetries.
 The leptonic charge corresponding to $l$ is split into three
 generation parts $l_i$. Only the $SU(2)$ anomaly cancelation condition
 changes~: $ L + n = 0 $ with $ L=\sum l_i$.
  The ``automatic'' requirement $n\neq 0$ leads to
  $N=3r$, $n=r$ and $L=-r$.
   So generation dependency lowers the smallest $N$ which admits
   automatic PQ to $N=6$. The potentials for $N=6$ are our $N=18$
   potentials from the generation independent case. As in the generation
   independent case $H\bar{H}A_1^2$  is forbidden for $N=3^i$.
   Since $n$ is not
   necessarily a multiple of 3 any more the operator classification
   for the term $H\bar{H}A_1^2$ is not valid. Only if $N=9r$
   this is still satisfied and general conclusions can be made~:
   There exist two types of (mod 3)--combinations of the $l_i$~:\\
   generation independent~:
   $(l_1,l_2,l_3)\; {\rm mod}\; 3=(0,0,0),(1,1,1),(2,2,2)$\\
   generation dependent~:$(l_1,l_2,l_3)\; {\rm mod}\; 3=(1,2,0)+$perm.\\
   Apart from $Q^3L$ and $u^2de$ which were forbidden
   in the generation independent case table~\ref{tab : op_patterns}
   remains valid for the B-violating operators. The L-violating operators
   with only one leptonic family index are allowed for $m+l_i$ mod $3=0$.
   Eq.~\refe{k_values} gives the corresponding values of $k$.
   Consequently for the second type only one neutrino flavor can receive
   ``large'' masses by one of the mechanisms discussed.
   Interestingly this generation dependent symmetries differ from
   the so called type II symmetries
   (i.~e.~$l_1=l_3\neq l_2$)~\cite{kap_mayr_nilles} necessary to
   solve the solar neutrino problem by allowing $L_{1,3}Qd$.\\

In conclusion we discussed axionic extensions of the SSM with a discrete
gauge symmetry to stabilize PQ-, B- and L-invariance. Our requirement
of an automatic PQ-symmetry excluded the Kim-type axion.
This led us to implementing the DFS-type axion and to solving
the $\mu$-problem.
We showed that a certain class of $Z_N$ symmetries with $N=9r$ can ensure
proton stability and simultaneously protect the axion solution.
To suppress the effect from higher dimensional operators a  small
tree-level axino mass ($<O(10^{-4})$ eV) is favored.
For a $\mu$-term $H\bar{H}A_1^2$ we found three classes of
discrete symmetries which show characteristic patterns of B- and
L-violating operators. One pattern of them allows only L-violating operators

For this class two sources of neutrino masses,
namely $L\bar{H}AA$ and $L\bar{H}N$, were discussed. They represent different
types of neutrino-axion connections with RHN scales $m_{3/2}$ and $f_a$,
respectively. If $L\bar{H}AA$ is present in general one neutrino mass
is fixed by the mass of the axino.
This excludes an axino mass $O($keV) resulting in a keV-Majorana neutrino.
An axino mass of $O(m_{3/2})$ can generate a favorable
neutrino mass of $O(10^{-3})$ eV though the axino solution is in danger.
On the other hand a low axino mass leads to a keV-Dirac neutrino.
Therefore we preferred the coupling $L\bar{H}N$.
A term $AN^2$ in the superpotential then can produce a RHN mass
$m_N \sim  f_a$ leading to Majorana neutrinos with mass of $O($eV).
The axino receives a one loop radiative mass $\sim m_{3/2}$ which contrary
to the tree level mass does not effect the  axion solution.
One model with this aspect was constructed for $Z_{36}$.

For the two simplest symmetries $Z_9$ and $Z_{18}$ we discussed all
acceptable superpotentials with less than five PQ-fields $A_i$. The $Z_9$
symmetry turned out to be incompatible with our requirements. In the
$Z_{18}$ case we gave examples for each of the three operator patterns and the
above aspects of neutrino-axion connections.\\

{\bf Acknowledgment}\\
We thank D. Kapetanakis and H. P. Nilles for discussions.

\end{document}